\documentclass[11pt, leqno, twoside]{article}
\usepackage{latexsym}   \usepackage{amsmath}    \usepackage{amsfonts}
\usepackage{amssymb}    \usepackage{amscd}
\newcommand{\cc} {{\cal C}} \newcommand{\ca} {{\cal A}}
 
\newcommand{\ch} {{\cal H}} \newcommand{\ce} {{\cal E}}

 \newcommand{\Ga} {{\mit\Gamma}}

\newcommand{\La} {{\mit\Lambda}} 
 \newcommand{\Om} {{\varOmega}}

\newcommand{\bn} {{\mathbb N}} 
 \newcommand{\bc} {{\mathbb C}}
\newcommand{\br} {{\mathbb R}} 
 
\newcommand{\lto} {{\longrightarrow}}
\newcommand{\rl} {{\;\rule{1.2ex}{1.2ex}\;\;}}
\newcommand{\bino} {{\bigskip \noindent}}
\newcommand{\bi} {{\bigskip}}
\newcommand{\bmp}{\begin{minipage}{12cm}}
\newcommand{\emp}{\end{minipage}}
\newcommand{\bmi}{\begin{minipage}{10.30cm}}
\newcommand{\emi}{\end{minipage}}

\begin{document}

\title{\bf Geometry and physics of today}

\author{{\bf Anastasios Mallios}}

\date{}

\maketitle

\pagestyle{myheadings} \markboth{\centerline {\small {\sc
{Anastasios Mallios}}}}
         {\centerline {\small {\sc {Geometry and Physics of Today}}}}

\begin{abstract}
The ``{\it geometry}'', in the sense of the classical differential
geometry of smooth manifolds (CDG), is put under scrutiny from the
point of view of {\it Abstract Differential Geometry} (ADG), along
with resulting, thereby, potential physical consequences, in what,
in particular, concerns physical ``{\it gauge theories}'', when
the latter are viewed as being, anyway, of a ``{\it geometrical
character}''. Yet, ``{\it physical geometry}'', in connection with
{\it physical laws} and the associated with them, within the
context of ADG, {\it ``differential'' equations} (whence, no
background spacetime manifold is needed thereat), are also under
discussion.
\end{abstract}

\setcounter{section}{-1}

\vskip 0.3in

$ $\hfill
\begin{minipage}{6cm}
    ``$\alpha\epsilon\acute{\iota}~o~\theta\epsilon\acute{o}\varsigma~\gamma\epsilon\omega\mu\epsilon\tau\rho\epsilon\tilde{\iota}$''

    (:\emph{``eternally the God geometrizes''})
\end{minipage}

\bi \bi

{\bf 1.}\quad By looking at the previous famous utterance
(attributed to Plato, according to Plutarch, see e.g. D.E. Smith
[31: p. 88, ft. 4]) as in the above frontispiece, while taking
also into account our nowadays conception of Physics, we can say
that;

\bino (1.1)\hfill
\begin{minipage}{12cm}
    \emph{``physical geometry''} is the \emph{outcome of} the
    \emph{physical laws}.
\end{minipage}

\bino In this regard, one might also refer here, for instance,
still to M.\;Faraday, as he is quoted by H.\;Weyl [35: p. 169], in
that [emphasis below is ours]:

\bino (1.2)\hfill \bmi
    ``... not the field should derive its meaning through its
    association with matter, but, conversely, ... \emph{particles
    of matter are ... singularities of the field}.''
\emi

\bino Now, by looking at the technical correspondence/association,
\begin{align*}
    \text{{\it physical law} $\longleftrightarrow$
    $\ca$-{\it connection},}\tag{1.3}
\end{align*}
one realizes that (1.1) might also be construed, as an
\emph{equivalent analogue of} the implication;
\begin{align*}
    &\text{$\ca$-\emph{connection} (:\;\emph{physical law})
    $\Longrightarrow$ \emph{curvature}}\\
    & \text{(:\,\emph{``geometry''}, alias,
    \emph{``shaping''}).}\tag{1.4}
\end{align*}
Consequently, still to repeat (1.1), thus, said it otherwise, one
concludes that;

\bino (1.5)\hfill
 \bmp
    it is actually the \emph{physical laws}, that \emph{make},
    what we might call (physical) \emph{``geometry''}.
 \emp

\bino Of course, we take for granted, concerning the above
terminology, the meaning of the technical term,
\emph{``$\ca$-connection''}, for which we refer thus, for
instance, to A.\;Mallios [12], or even to [13], [17].

Now, it is worthwhile to comment here, a bit more, on the inverted
commas, put above on the word,
\begin{align*}
    \text{\emph{geometry}.}\tag{1.6}
\end{align*}
Indeed, the same are meant, as well as, hint, therein, at the
technical and also fundamental, in point of fact, issue, which the
aforesaid (Greek) word contains in itself; \emph{namely}, the
\emph{entanglement of ourselves}, in that point of view, or even,
the manner we look at that notion, as this is
implemented/understood, exactly, by the second component of the
same word (the latter being, in effect, a \emph{composed one},
that is, the Greek verb, \emph{``metr\={o}''} (:\;measure).
Accordingly, any time we refer to/use that notion, \emph{by
definition}, viz. by the real essence of the same word,

\bino (1.7)\hfill
 \bmp
    the term \emph{``geometry''} does not actually correspond
    to/means something physical (:\;\emph{real}), but, simply,
    \emph{a model} of ours, \emph{pertaining to} the description
    of \emph{reality} (in whatever sense of the latter concept).
 \emp

\bino Furthermore, it is still appropriate to remind us, at this
point, of A.\;Einstein's maxim, in that;

\bino (1.8)\hfill
 \bmp
    \emph{``Time and space are modes by which we think, not
    conditions in which we live''}.
 \emp

\bino See thus, for instance, Yu.I.\;Manin [24: p. 71], as well
as, within actually the same vein of ideas, (1.29) in the sequel.
[Emphasis in (1.8) above is ours, as it will also be the case,
occasionally, in quotations, throughout the sequel]. Yet, we
mention here the relevant remarks of P.G.\;Bergmann [3: p. 33], in
that,

\bino (1.9)\hfill
 \bmp
    ``Einstein ... \emph{did not consider geometrization of
    physics} a foremost or even \emph{a meaningful objective...''}
 \emp

\bino (I am indebted here to I.\;Raptis for bringing to my
attention the previous citation of Bergmann). Yet, the same
author, as above (loc. cit.), insists in that, what is \emph{of
importance} here \emph{is},

\bino (1.10)\hfill
 \bmp
    ``... \emph{not a geometric formulation or picturization but a
    ... fusing of} the \emph{mathematical structures intended to
    represent physical fields}.''
 \emp

\bino We remark here that the above are still in accord with (1.1)
or (1.5) in the preceding. Thus, we are led again, herewith, to a

\bino (1.11)\hfill
 \bmp
    \emph{``relational aspect''} of what we might call,
    \emph{``physical geometry''}.
 \emp

\bino In other words, we thus arrive at \emph{something}, which is
\emph{more close to what}, as we still mentioned above, \emph{we}
have already \emph{said by} (1.5). Furthermore, this same aspect
is also \emph{akin to what we} may \emph{understand}, as we shall
see later on, when speaking of

\bino (1.12)\hfill
 \bmp
    \emph{``geometry''}, determined \emph{by ``differential''
    equations}, yet, the \emph{``solution space''} of the latter.
    The same might still be conceived, even, as the
    \emph{source}(!)\;\,of the \emph{``cartesian point of view''};
    however, \emph{see} also \emph{(1.14)} in the sequel,
    concerning  that perspective, within the present
    \emph{abstract} (thus, \emph{space-independent}\,(!))
    \emph{setting}.
 \emp

\bino So, still, within the aforesaid context (see also e.g.
(1.11)), we can further say that;

\bino (1.13)\hfill
 \bmp
    \emph{``geometrization'' of physics} means, in point of fact,
    \emph{``arithmetization''} of the same, for our
    \emph{``geometry''} is, in effect, \emph{``arithmetical''},
    that is, \emph{``cartesian''}(!), in character, hence,
    \emph{not} a \emph{physical} (:\;natural) one!
 \emp

\bino Consequently, one comes to realize that,

\bino (1.14)\hfill
 \bmp
    \emph{the} previous \emph{association} becomes thus \emph{more
    natural}, to the extent that it is \emph{more
    ``relational''}(!), in nature.
 \emp

\bino However, what is also here of a particular significance,
concerning the whole subject matter of the present work, the
preceding point of view, as in (1.12), is actually meant in an

\bino (1.15)\hfill
 \bmp
    entirely \emph{``space independent''} way, that is, \emph{not}
    in a \emph{``cartesian-wise''} manner,
 \emp

\bino as this also will become clear, along with the terminology
applied herewith, through the subsequent discussion. That is, in
other words, based on the \emph{abstract formalism} of the same
technique of \emph{Abstract Differential Geometry} (ADG), one is
able to

\bino (1.16)\hfill
 \bmp
    formulate \emph{``differential'' equations without} having the
    need to resort to any \emph{background}
    (\emph{``cartesian''--``newtonian''}, so to say)
    \emph{``space''}, to work with.
 \emp

\bino This latter situation might be, in point of fact, as we
shall see in the sequel, of paramount significance for problems of
\emph{quantum gravity}, when the same problems are viewed from the
standard perspective, viz. from that one of the classical
differential geometry of smooth (:\;$\cc^\infty$-)manifolds (CDG).

So, in accordance with (1.11), one gets, indeed, at a
\emph{``leibnizian''}, so to say, \emph{point of view}, that is,
by following Leibniz himself,

\bino (1.17)\hfill
 \bmp
    we should find a \emph{``geometrical calculus''} that operates
    directly on the \emph{``geometrical objects''} without the
    intervention of coordinates.
 \emp

\bino In this regard, we may even remark here, anyway, that
\emph{the latter function}, as above (:\;coordinates) \emph{is},
for that matter,
\begin{align*}
    \textit{``... an act of violence''.}\tag{1.18}
\end{align*}
See thus H.\;Weyl [36: 90]. On the other hand, concerning (1.17),
cf., for instance, N.\;Bourbaki [5: Chapt. I; p. 161, ft. 1].
Furthermore, within the same context, one has here the relevant
remarks of B.\;Riemann, in that;

\bino (1.19)\hfill
 \bmp
    ``Specifications of mass [:\;measurements] require an
    \emph{independence of quantity from position}, which can
    happen in more than one way''.
 \emp

\bino Cf., for example, A.\;Mallios [14: (1.3)]. Thus, in toto,
the preceding sustain, indeed, the aspect that:

\bino (1.20)\hfill
 \bmp
    the description of the \emph{physical laws}, something that
    could also include the \emph{quantum r\'{e}gime}, as well,
    should be made in such a manner, that \emph{no supporting
    space}, or even \emph{space scaffolding} (:\;framework),
    essentially contributing to that description, \emph{is to be
    included in our ``calculations''} (:\;rationale); hence,
    \emph{the latter have} thus \emph{to be} entirely
    \emph{independent of any notion of ``space''} of the aforesaid
    type.
 \emp

\bino Now, the previous aspect of \emph{``description of physical
laws''}, as in (1.20), can, in point of fact, be conceived, as
just referring to the very \emph{``geometrical calculus''} \`{a}
la Leibniz (cf. (1.17)), hence, to this same
\emph{``geometry''}(!), in that respect, in the sense of Leibniz,
or even, to its \emph{``relational point of view''}, according to
(1.11). Furthermore, the same perspective of

\bino (1.21)\hfill
 \bmp
    \emph{``geometry''}, as \emph{``description}(study) \emph{of
    physical laws''},
 \emp

\bino leads, of course, simply, to the aspect of,

\bino (1.22)\hfill
 \bmp
    doing \emph{``geometry'', via ``differential'' equations} (a
    fact that actually goes back to Ren\'{e} Descartes himself:
    \emph{``Analytic Geometry''}),
 \emp

\bino as exactly hinted at, already, by (1.12) in the foregoing.
Now, we are just going to comment further on the latter aspect, as
appeared, within the present abstract setup of ADG,
straightforwardly, by the next Section, making thus also still,
more clear, our previous remarks in (1.16), as above.

\bi \bi
 {\bf 2.\quad ``Differential'' equations in the setting of ADG.
Functoriality.}$-$ As already mentioned above, our aim, by the
following discussion, is virtually to \emph{clarify} (1.16),
\emph{and} to look also at further \emph{consequences} thereat:

Thus, to start with, we can certainly remark that, one of the most
effective methods, thus far, of describing \emph{physical laws}
has been, of course, that one, provided by \emph{``differential
equations''}; hence, the foremost \emph{applications} thereof
\emph{of} the (classical) \emph{differential geometry} (CDG,
indeed, \emph{Calculus}(!), yet, of \emph{``the glittering
trappings of Analysis''}, to recall here G.D.\;Birkhoff; see, for
instance, A.\;Weinstein [34:\;p.1, ft.2]).

However, the latter (viz. the \emph{classical}) way of describing
physical laws contains in itself, already, the seeds of the
defaults, that exactly should be avoided, just by virtue of our
previous remarks, as in (1.20). Indeed, by the very characters of
\emph{the classical theory} (:\;CDG), \emph{its whole machinery}
(\emph{mechanism}) \emph{is} entirely \emph{rooted on the
supporting space} (viz. on the \emph{``locally euclidean''} smooth
manifold). Accordingly, simply, as a result of (1.20), one
concludes that;

\bino (2.1)\hfill
 \bmp
    \emph{the notion of a} (locally
    euclidean--smooth--)\emph{manifold} proves thus \emph{not} to be
    the \emph{appropriate} one(!), in order \emph{to describe
    physical laws} (:\;the \emph{``reality''}) to the extent,
    \emph{at least}, that the latter refer \emph{to the quantum
    deep}, as well.
 \emp

\bino In this context, we may still recall herewith, the relevant
comments of A.\;Einstein himself, pertaining, in point of fact, to
the

\bino (2.2)\hfill
 \bmp
    \emph{inappropriateness of the manifold concept for physical
    reality}(!).
 \emp

\bino See, for instance, A.\;Mallios [16: (1.6)]. On the other
hand, one can further say that,

\bino (2.3)\hfill
 \bmp
    \emph{the} aforementioned \emph{drawback of the notion of
    smooth manifold in} problems connected with \emph{the quantum
    deep is} mainly \emph{due, not only}\,(!) \emph{to the way, we
    consider arising} the \emph{``differential-geometric''
    mechanism}, within the context of CDG (see thus, however,
    (3.3) in the sequel), but,

    \bino (2.3.1)\hfill
    \bmi
        \emph{much more}, because \emph{we} still \emph{keep, as a
        ``working framework'', the whole ``space''}, viz. the
        entire smooth manifold itself,
    \emi

    \bino by further looking at it, even \emph{locally}, as domain
    of definition of what \emph{we} define, as
    \emph{``differentiable functions''}.
 \emp

\bino Now, the latter point of view, as mentioned in (2.3) above,
proves to be, by concrete working examples we present below, a
quite \emph{unnatural way} of trying to apply the
\emph{``differential geometric mechanism''} of CDG, its
\emph{character} being, in point of fact, entirely
\emph{algebraic}(!), as we are still going to clarify in the
sequel. Furthermore, it is this same aspect, as above, where we
are usually confronted with an \emph{extremely pestilential
anomaly} of the classical theory, \emph{pertaining}, in
particular, \emph{to the quantum deep}, this being thus, indeed,
the \emph{main source of ``infinities''}
(:\;\emph{``singularities''})! Notwithstanding, all these
anomalies, \emph{without} actually \emph{being real} ones\,(!)
(cf. thus the aforementioned examples, as presented by the ensuing
discussion).

On the other hand, we further illuminate \emph{the situation that
appears, within the quantum framework}, when looking at it
\emph{from the point of view of the abstract theory}, summarizing
thus briefly the relevant conclusions into the following.

\bino {\bf Scholium 2.1.}$-$ When looking at the fundamental of
\emph{quantum theory}, in conjunction with potential applications
in that context of (differential) geometry, one actually realizes
that;

\bino (2.4)\hfill
 \bmp
    we usually \emph{associate numbers} (\`{a} la Descartes)
    \emph{to a space that}, in effect, \emph{does not exist}(!),
    in the sense, at least, \emph{we} ascribe to it, that
    \emph{``spatial perspective''} of ours being, in point of
    fact, \emph{always cartesian}(!), something, of course, which
    is \emph{not in accord with} our (experimental) knowledge, as
    it concerns \emph{the quantum r\'{e}gime}.
\emp

\bino Thus, we are, indeed, trapped here, by our own perspective,
due actually to our preexistent assumption, pertaining, as a
matter of fact, to the manner we consider our \emph{``calculus''}
(hence, of course, that same \emph{instrumental issue of}
(classical) \emph{differential geometry}, as well) is virtually
arrived, this being thus, according to the classical theory,
\emph{``locally euclidean''}, viz. \emph{``newtonian''}, in nature
(:\;manifold$\;\leftrightarrow \;$spacetime). Therefore, as a
consequence, we are thus

\bino (2.5)\hfill \bmp
    \emph{unable to apply} the classical (:\;newtonian) aspect of
    \emph{differential geometry in the ``quantum deep''}, due
    mainly to the emergence of the so-called
    \emph{``singularities''}, and other relevant anomalies. [As
    already said, several times in the preceding, the latter
    phenomenon being actually due to the \emph{particular type of}
    our (\emph{``smooth''}) \emph{functions involved}, that
    \emph{``smoothness'' being}, in turn, \emph{a direct outcome}
    of the sort \emph{of ``space''} (:\;locally euclidean)
    \emph{we use}!].
\emp

\bino Consequently, \emph{once more},

\bino (2.6)\hfill \bmp
    \emph{it is not the functions} we use (viz., when considering
    them, as carriers of the \emph{``differential
    geometric-mechanism''}, from the point of view of ADG, that
    is, so to say, in the \emph{``leibnizian'' perspective} of the
    latter term), which are \emph{inappropriate, concerning the
    quantum deep, but}, simply, \emph{the ``space''}, on which
    \emph{the} said \emph{functions are supposed to be defined},
    such \emph{a space}, as that one we try to apply (viz. the
    ``locally euclidean'' one), \emph{being} virtually
    \emph{non-existent}(!), in that context, and \emph{not only
    this},
\emp

\bino given that,

\bino $ $\hfill \bmp an \emph{``arithmetical space''}, as it
actually is the
    standard \emph{``eu\-cli\-dean\-/car\-tesian space''}, which we
    usually employ in the \emph{classical theory} (:\;CDG), is
    not, of course, \emph{``physical''} (:\;\emph{real})(!), as
    this is, in effect, realized when, in particular, referring to
    the \emph{``quantum deep''}. [We thus get, in that\linebreak
\emp

\bino (2.7)\hfill
 \bmp
   context,
    even an \emph{``experimental''} (:\;concrete)
    \emph{ascertainment} of the \emph{ineffectiveness of} our
    (spatial) \emph{model}!]. That is, the \emph{``space''} model,
    we usually ascribe in our physical theories, to what we
    actually understand, as \emph{``physical space''}, is entirely
    a \emph{numerical one''}. In that context, we are influenced, of
    course, from our own successes, so far, in the macroscopic world.
    This model, however, collapses when confronted with the
    quantum deep.
 \emp

\bino So, in other words, we are thus entrapped, in that respect,
by the \emph{particular success}, thus far, of the aforesaid point
of view (:\;the classical one), in what, namely, especially
concerns our experience/applications \emph{``in the large''}.

Now, within the same vein of ideas, and still, in connection with
the (\emph{categorical}), in effect, \emph{correspondence} amongst
\emph{``space''} and \emph{functions}, one actually has, in that
respect, the following \emph{``identification''},
\begin{align*}
    \textit{functions $\rightleftarrows$ space, ``Gel'fand
    duality''}\tag{2.8}
\end{align*}
which we may also call (already depicted above), \emph{Gel'fand
duality''}, a fact strikingly pointed out, in its full generality,
by the language of the \emph{theory of} (especially,
\emph{non-normed}) \emph{Topological Algebras} (see, for instance,
A.\;Mallios [TA; p. 223, Theorem 1.2, as well as, p. 227, Theorem
2.1]).

\bino Now, as already hinted at in the foregoing, and which will
also be considered, by the ensuing discussion, the previous
situation, has nothing to do, in effect, with the \emph{mechanism
itself} of the aforesaid classical theory (:\;differential
geometry), the same machinery being essentially
\emph{``leibnizian''}(!), in nature, as this, indeed, has been
pointed out, by what we may call \emph{``Abstract Differential
Geometry''} (:\;ADG); see thus A.\;Mallios [13], as well as, [17].

In toto, the preceding represent the way one may look at what we
usually understand, nowadays, as \emph{``space''} (speaking, of
course, in terms, of what we call \emph{``mathematical
physics''}). True, the previous thoughts are actually the
\emph{outcome} of our experience derived \emph{from} ADG, while
the same still supplies potential applications in problems of
\emph{quantum} relativity, as the latter has been explained
already in other places (see, for instance, A.\;Mallios [17], as
well as, A.\;Mallios--I.\;Raptis [20], [21]). So it is this
entirely \emph{new} (\emph{axiomatic}) \emph{perspective of} ADG,
pertaining to the  \emph{inherent mechanism of} the classical
\emph{differential geometry} (:\;CDG), which provides several
potential applications, while the same mechanism proves, very
likely, to be also in accord with the \emph{``spatial''
situation}, one is confronted with \emph{in the quantum deep}, as
already hinted at, by the foregoing discussion; in this regard,
see also e.g. A.\;Mallios--E.E.\;Rosinger [23], along with
A.\;Mallios--I.\;Raptis [21]. On this latter aspect we are still
going to present, however, some further illuminating comments,
through the subsequent discussion, as well.

On the other hand, by looking at the whole \emph{classical set-up
from the point of view of} ADG, we can still point out here that,
by \emph{complete contrast with} the situation, which usually
dominates \emph{the classical case},

\bino (2.9)\hfill
 \bmp
    \emph{the framework of} ADG \emph{does not}, in principle,
    \emph{depend on any background ``space''} (:\;carrier, think
    e.g. of \emph{``space-time''} for the classical domain), that
    would contribute to its \emph{``differential''} equipment, the
    latter being thus \emph{entirely rooted on} $\ca$(!), our
    \emph{``generalized arithmetics''}, alias, \emph{``sheaf of
    coefficients''}.
 \emp

\bino Yet, the latter issue in (2.6), as above, constitutes, in
point of fact, still,

\bino (2.10)\hfill
 \bmp
    the quintessence of the \emph{quantum field-theoretic
    character of} ADG.
 \emp

\bino Indeed, \emph{the} whole \emph{set-up of} ADG
\emph{becomes}, by its very definition, \emph{susceptible of}

\bino (2.11)\hfill
 \bmp
    \emph{formulating} our \emph{equations in} a \emph{quantum
    field-theoretic manner},
 \emp

\bino viz. \emph{quantum-relativistically}! In this connection,
see also our previous relevant remarks in A.\;Mallios [14: (9.8),
(9.23), along with Section 11 therein]. Yet, to put the above
subject matter still in an equivalent way, we can further remark
here that;

\bino (2.12)\hfill
 \bmp
    \emph{it is} actually \emph{we, who describe the} (physical)
    \emph{laws}, read, \emph{``differential'' equations}, by means
    of our \emph{``arithmetics''}, thus, for the case at issue,
    through \emph{the} ($\bc$-algebra) \emph{sheaf} $\ca$,
    while the same machinery (:\;\emph{calculus''}, \`{a} la
    Leibniz, or even \emph{``differential geometry''}) is still
    \emph{based on} $\ca$, and not on any background
    \emph{``space''}, at all(!), as it was classically the case.
    Of course, this latter fact may be of \emph{paramount
    importance}, when one is confronted with problems of
    \emph{quantum gravity}.
 \emp

\bino In this regard, one may still refer to the relevant comments
of J.\;Baez [2: beginning of Preface], in that (emphasis below is
ours):

\bino (2.13)\hfill
 \bmp
    ``A fundamental problem with \emph{quantum ... gravity} ... is
    that \emph{in ... general relativity} there is no background
    geometry to work with: \emph{the geometry} of spacetime itself
    \emph{becomes a dynamical variable''.}
 \emp

\bino On the other hand, the aforementioned (see (2.9), (2.12))

\bino (2.14)\hfill
 \bmp
    \emph{independence of} the \emph{``differential'' mechanism
    of} ADG \emph{from any background space}, gives to that
    mechanism the possibility to be considered, as, a
    \emph{``variable''} entity too, the same \emph{being}, by its
    very construction, entirely \emph{based on} (reduced to)
    $\ca$; therefore, what we also understand, as
    \emph{``differential'' geometry} (:\;\emph{``geometrical
    calculus''}, \`{a} la Leibniz), entailed thereof,
    \emph{becomes} still a \emph{``variable''}, as well.
 \emp

\bino Furthermore, that also appears \emph{fundamental}, herewith,
the same \emph{``geometrical calculus''}, hence, the concomitant
\emph{``geometry''} too, becomes simply \emph{``relational''},
referring thus directly to the \emph{``geometrical objects''} (in
our case, \emph{vector sheaves}) themselves, without the
interference, of any \emph{``space''}, in the classical sense of
the latter term.

In this connection, we also recall, for convenience,
\emph{technically speaking}, that: We suppose herewith that we are
thinking, in terms of an (\emph{abstract}) \emph{``differential
setting''}, based on a given \emph{``differential triad''},
\begin{align*}
    (\ca, \partial, \Om )
    \tag{2.15}
\end{align*}
over an (arbitrary, in general) \emph{topological space} $X$, base
space of all the sheaves involved, throughout. Now, within that
context, a \emph{``geometrical object''}, thus, for instance, an
elementary particle, can be associated with what we call a
\emph{Yang-Mills field}, viz. a pair
\begin{align*}
    (\ce, D),\tag{2.16}
\end{align*}
consisting of a \emph{vector sheaf} $\ce$ on $X$ and an
$\ca$-\emph{connection $D$ on} $\ce$; see e.g. A.\;Mallios [16:
(3.2), (3.3)], or even [17: Chapt. VI]. It is actually in terms of
such pairs, as above, that \emph{``differential'' equations}, in
the framework of ADG are referred (loc. cit.).

Thus, within the above set-up, we can further refer here to a
\emph{fundamental principle}, in effect, of the whole machinery,
thus far, \emph{of Abstract Differential Geometry} (:\;ADG), in
that;

\bino (2.17)\hfill
 \bmp
    \emph{everything}, that \emph{we want to ascribe to} a pair
    $(\ce ,D)$, as above, \emph{is} virtually \emph{reduced to} a
    similar condition/asumption for \emph{the pair} $(\ca,
    \partial)$, see (2.15), yet, occasionally, under appropriate (in
    principle, \emph{only}(!)) \emph{topological} hypotheses for
    $X$ (see also the subsequent comments).
 \emp

\bino As already noted before, the context of (2.17) exhibits, in
point of fact, the \emph{``Leitmotiv''} that actually dominates
the very technique of ADG; see thus A.\;Mallios [VS], or even
[17]. On the other hand, the same ensures also the

\bino (2.18)\hfill
 \bmp
    \emph{``covariance'' of} the whole setting of ADG, \emph{with
    respect to} $\ca$.
 \emp

\bino Thus, the \emph{``variance''} here is always \emph{relative
to our} own \emph{``arithmetics''}, or even (\emph{generalized})
\emph{domain of coefficients''}, yet, \emph{``structure sheaf''}
$\ca$ (by assumption, a \emph{unital commutative $\bc$-algebra
sheaf} on $X$, cf. (2.15)), which, for that matter, is, of course,
the case, as well; so, strictly speaking, it is actually \emph{we}
always, who \emph{measure}(!)/\emph{calculate}, while, and this is
also \emph{of} a particular \emph{importance}, as already pointed
out in the preceding, this whole framework/calculations of ADG,
\emph{without} actually \emph{leaning upon any background
``space''} (:\;carrier), as, for instance, ``space-time''(!) of
the classical case.

On the other hand, we can further say that;
\begin{align*}
    \text{\emph{physical laws are} always
    \emph{``functorial''}.}\tag{2.19}
\end{align*}
Of course, in point of fact, we \emph{``abuse language''} here,
when referring to the above statement, as explained by the
following.

\bino $ $\hfill
 \bmp
    \hspace{0.5cm}{\bf Note 2.1.}$-$ Looking at the sense, we
    actually use the term,
    \[
        \textit{``functorial''}
    \]
    as in (2.19) above, and also in conjunction with (1.2) in the
    preceding, we should still remark here that;

        \bino (2.20)\hfill
    \bmi
    \emph{the aforesaid term is always meant, with respect}, in
    effect, \emph{to us}(!), viz. relative to (our
    \emph{``generalized arithmetics''}) $\ca$.
    \emi
 \emp

 $ $\hfill
 \bmp
   Therefore, what we actually consider in (2.19) is, in point of
    fact, the \emph{manifestation of the physical laws}!
 \emp

\bino Now, this goes, of course, hand in hand, with (or even, it
is something that is, in point of fact, an \emph{equivalent
expression} of) the \emph{``principle of general covariance''}.
Accordingly, by considering now, as we did it in the preceding,
\emph{differential equations, as expressing physical laws} (see
(1.5), (1.12)), we can realize that, indeed,

\bino (2.21)\hfill
 \bmp
    \emph{differential equations} should be, by their very
    definitions, \emph{``functorial'', in nature}! Consequently,
    their formulation should be made, \emph{in terms of
    ``functorial objects''}, as well.
 \emp

\bino Now, by the last term, when speaking in technical language,
we mean, of course, something that, by definition, is
$\ca$-\emph{invariant}, alias a \emph{``tensor''}, in the sense
that it respects our \emph{``arithmetics''} $\ca$.

Furthermore, (2.19) can still be construed, as an outcome of
(1.2), in conjunction with (1.5) in the preceding. Thus, by
further considering (see also (2.20), as above) the

\bino (2.22)\hfill
 \bmp
    \emph{physical laws}, as \emph{the manifestation of} the
    (deepest physical) \emph{dynamics} (:\;\emph{``causality''}),
    one comes to the conclusion that;

    \bino (2.22.1)\hfill
    \bmi
     \emph{``dynamics''} should be \emph{``functorial''}, as well,
    \emi

    \bino whenever \emph{we} actually effectuate it (viz. the
    physical law, cf. also (1.4)). Therefore, this very
    realization of it (\emph{by us}(!), of course) becomes
    \emph{``functorial''}, or even \emph{``tensorial''} too,
    hence, the same physis of the \emph{curvature}
    (:\;\emph{``geometry''}), see also (1.4), as before.
 \emp

\bino Now, by further commenting on our last conclusion, as above,
we still recall that, according to our axiomatics,

\bino (2.23)\hfill
 \bmp
    the \emph{curvature} (:\;field strength) is the manifestation
    (effectuation) of the \emph{``identification''}
    (correspondence, cf. also (1.4)),
    \[
        \text{\emph{dynamics} (:\;``causality'')
        $\longleftrightarrow$ ($\ca$-)\emph{connection}},
        \tag{2.23.1}
    \]
    therefore (see also (2.22.1)), the \emph{tensorial}
    (\emph{functorial}, cf. (2.20)) \emph{aspect of} the
    \emph{curvature}.
 \emp

\bino In this connection, we can still note that the
aforementioned \emph{functorial/tensorial character of the
curvature}, in the sense, of course, of (2.20), being always
\emph{the outcome} (field strength) \emph{of a} given
\emph{``field''} (:\;$\ca$-connection, see, for instance,
A.\;Mallios [16: (3.15)] or even [14: (3.21.1)]) \emph{is} further
\emph{expressed}, by the familiar relation,
\[
    \nabla \rho =0,
    \tag{2.24}
\]
yet, \emph{equivalently} (precisely speaking, in terms of the
formalism of ADG), by the relation;
\[
    D_{\ch om (\ce ,\ce^* )}(\tilde{\rho})=0,
    \tag{2.24}
\]
where we still have;
\[
    \ch om (\ce, \ce^* )= \ce^* \otimes_\ca \ce^* =(\ce
    \otimes_\ca \ce )^* .
    \tag{2.26}
\]
See A.\;Mallios [VS: Chapt. VII; p. 165, (8.70), along with Chapt.
IV; p. 302, The\-o\-rem 6.1 and p. 305: (6.16)]; thus, we have
herewith the so-called, classically, \emph{``Levi-Civita
identity''}. By further referring to the above notation, we
consider therein a given \emph{Yang-Mills field}
\[
    (\ce, D),
    \tag{2.27}
\]
see loc. cit., Chapt. IX; p. 244, along with (2.15), as above,
while $\rho$ stands there for a \emph{Riemannian $\ca$-metric} on
$\ce$, \emph{``compatible with $D$''} (ibid., Chapt. VII; Section
8). It is worth noticing here that the previous condition on the
pair
\[
    (D,\rho),
    \tag{2.28}
\]
as above, is actually the \emph{upshot of a similar assumption for
the} standard \emph{pair},
\[
    (\ca, \partial),
    \tag{2.29}
\]
cf. (2.14), under appropriate supplementary conditions on the
items involved herewith, these being in the case of $X$,
\emph{only topological} ones (cf. thus (2.16) in the preceding);
yet, in that context, see also A.\;Mallios [VS: Chapt. VII; p.
168, Theorem 9.1: \emph{Fundamental lemma of Riemannian vector
sheaves}]. Accordingly, we further understands here that (:
\emph{``physical significance'' of} (2.25)),

\bino (2.30)\hfill
 \bmp
    \emph{to ``realize'' the curvature}, one has to
    \emph{``compare'' it with something else}!
 \emp

\bino We terminate the present Section with the subsequent remarks
of N.\;Bohr, as quoted e.g., by S.Y.\;Auyang [1: p. 229]),
referring to the way one actually has to look at the Nature;
indeed, with the same remarks the foregoing rationale and related
remarks thereon are really in accord, as it actually concerns
\emph{our relevance}, with respect to the observed physical laws,
which, technically speaking, as it was pointed out in the
preceding, is \emph{expressed}, in effect, \emph{through the
``structure sheaf'' $\ca$, independently of any
surrounding}/supporting \emph{``space''}. Thus, according to the
aforementioned remarks (emphasis below is ours),

\bino (2.31)\hfill
 \bmp
    ``It is wrong to think that the task of physics is to point
    out how nature is. \emph{Physics concerns what we can say
    about nature}.''
 \emp

\bino Consequently, to follow in that context the favorite
parlance of A.\;Einstein himself, \emph{we} thus always
\emph{``describe''}, hence, \emph{not} \emph{explain}\,(!), the
\emph{physical applications} of every day life being, therefore,
simply, \emph{consequences of the former} (descriptions), as
above(!). Yet, within the same vein ideas, we may still quote,
herewith, L.\;Wittgenstein [38: p. 17], in that;

\bino (2.32)\hfill
 \bmp
    \emph{``Physics does not explain anything; it simply describes
    concomitant cases''}.
 \emp

\bino (Emphasis above is ours). Therefore, as already emphasized
in the preceding, we do not actually explain \emph{``anything''},
through Physics, as it concerns the \emph{physical laws}
(:\;physis), but, just, \emph{describe}/study \emph{their
consequence}(!); notwithstanding, as a consequence, however, of
the latter function, it undoubtedly seems (cf.
\emph{applications}) that,

\bino (2.33)\hfill
 \bmp
    \emph{we do understand,} nevertheless, several times and, of
    course, \emph{always},  \emph{to a certain extent}(!), \emph{the way} that
    \emph{these laws work}!
 \emp

\bi \bi {\bf 3.\quad ADG, as applicable in the quantum deep.}$-$
Our purpose, by the ensuing discussion, as the title of this
Section indicates, is to further clarify the way one can look at a
\emph{potential application of} ADG \emph{in the quantum
r\'{e}gime}, thus, in point of fact, of \emph{the very mechanism
of the classical} (:\;\emph{``newtonian''}) \emph{differential
geometry}, very effective(!), for that matter, so far, however,
\emph{now, within the aforesaid domain, but}, already \emph{from
the point of view of} ADG (viz. \emph{axiomatically}), thus,
\emph{freed from} its \emph{``beautiful shackles''} (C.J.\;Isham);
indeed, it is proved that the latter \emph{obstacles} are
\emph{due}, simply, \emph{to the entanglement}, according to the
classical theory, \emph{of the} same \emph{mechanism with the
``locally euclidean'' nature} of that theory (in effect,
\emph{much more}, because of the \emph{maintenance of the} whole
\emph{``smooth setting'', as} a \emph{working framework}, in this
context, cf. also (2.3) in the preceding), the latter being also
considered, in view of the same standard theory (CDG), \emph{the
only source}(!), within that context, \emph{of} the all powerful
(infinitesimal/integral) \emph{Calculus}, hence, of the classical
differential-geometric machinery, as well. So, it is here exactly
that a \emph{supreme didagma of} ADG comes just to the foreground,
\emph{in fact};

\bino (3.1)\hfill
 \bmp
    \emph{the differential-geometric mechanism of} the
    \emph{classical differential geometry} (CDG)--being, in effect,
    \emph{of} a \emph{leibnizian character--can}, equally well,
    \emph{be supplied, by other sources, apart from a ``locally
    euclidean'' space}/(smooth) manifold, its \emph{existence}
    being thus \emph{independent of any} such \emph{``space''}.
 \emp

\bino Furthermore, as already pointed out in the preceding (see,
for instance, the quoted citations, in that context, of Einstein,
Feynman, Isham), a \emph{``space''}, as \emph{in} the \emph{latter
part of (3.1)}, together \emph{with its ``differential set-up'',
is} entirely \emph{out of the question for the quantum deep}(!).

On the other hand, by further commenting, within the preceding
vein of ideas, on the basis of our experience from ADG, as exposed
above, we realize that one can virtually interrelate well-known
phenomena in the past with still existing tendencies in quantum
physics of today:

Thus, the heuristic \emph{opposition of Einstein against Quantum
Field Theory} (:\;\emph{``the other Einstein''}, see e.g.
J.\;Stachel [33: p. 283, 285]) might also be viewed, apart from
other physical reasons, still, as an outcome of the \emph{failure
of} classical differential geometry --hence, in particular, of
general relativity too-- as it concerns \emph{the way the inherent
in that theory} (differential) \emph{``Calculus'' is supplied, to
cope with problems of the quantum theory}. Indeed, we can further
say that, looking at the same classical (:\;\emph{``newtonian})
manner of definition of the \emph{``derivative''},

\bino (3.2)\hfill
 \bmp
    Einstein was demanding, within that framework, to \emph{abandon,
    even} the notion of \emph{continuity}(!) \emph{in physics},
    having thus, instead, to invent a \emph{``purely algebraic
    physics''} (loc. cit., p. 285); therefore, in particular, as,
    of course, we can say, a (purely) \emph{algebraic
    analysis}(!), as well.
 \emp

\bino In this connection, we can certainly refer here, as already
done in the preceding, to the relevant remarks thereof of
R.P.\;Feynman [7: p. 166]. as well as, to those of C.J.\;Isham [9:
p. 393] (in this regard, see also e.g. A.\;Mallios [16]),
concerning, namely, the \emph{ineffectiveness of } the
\emph{classical differential geometry}, accordingly, of that one
of a \emph{smooth} (:\;$\cc^\infty$-)\emph{manifold} too,
\emph{within the quantum r\'{e}gime} (yet, see (2.3) in the
foregoing, along with our discussion in the subsequent Section 4).

On the other hand, the \emph{pertinence}, in that context,
\emph{of} ADG \emph{to} confronting with problems of \emph{quantum
gravity} still lies in its \emph{algebraic} (viz.
\emph{``leibnizian''}, so to say) \emph{character}: Indeed, the
whole edifice of ADG is, by its very construction,
\emph{sheaf-theoretic, sheaf theory} being, of course, of an
\emph{algebraic nature} (see, for instance,
H.\;Grauert--R.\;Remmert [9: p. VII]). Thus, ADG \emph{might} also
\emph{be construed}, as an

\bino (3.3)\hfill
 \bmp
    \emph{algebraic} (:\;\emph{``leibnizian''}) \emph{manner of
    presenting} the fundamentals of the \emph{classical
    differential geometry}, while, at the same time, still
    getting, as an \emph{outstanding outcome} (see, for example,
    (2.22), as well as, (2.1) in the preceding), the
    \emph{possibility of working, without any resort to a
    background ``space''}, in the classical sense of the latter
    term, as for instance, to a \emph{``space-time
    continuum''}(!), as it happens, instead, in the standard
    theory.
 \emp

\bino Certainly, the significance of the aforementioned \emph{two
issues of} ADG cannot be underestimated, while the same might be,
in effect, quite well, what A.\;Einstein himself, by 1935 already,
was looking for (see, for instance, still, J.\;Stachel [33: p.
285], as above).

\bi \bi {\bf 4.\;Particular potential applications of ADG in the
quantum r\'{e}gime.}$-$ We start, by presenting, within the
framework of ADG, the relevant \emph{theory of Elem\'{e}r
E.\;Rosinger}, pertaining to \emph{``generalized functions''},
whose \emph{algebra} (\emph{sheaf}), in particular, the
\emph{``foamy''} one, can be used, as a \emph{``sheaf of
coefficients''}, defining thus, appropriately, a corresponding
herewith \emph{``differential triad''}, basic ingredient to having
a set-up in developing the mechanism of ADG (see (2.19) in the
foregoing). For similar previous accounts, see also
A.\;Mallios--E.E.\;Rosinger [22], [23], as well as, A.\;Mallios
[17: Chapt. IX; Section 5].

However, before we come to the relevant exposition, it is still to
be noticed, herewith, a \emph{fact of a particular significance},
referring to the very \emph{structure of} ADG (cf.\;(4.1) below,
along with Subsection 4.(b) in the sequel), as it concerns two
\emph{im\-po\-rtant special cases} of the general theory of ADG,
we are going to consider, by the subsequent discussion; the same
are also characteristic of the way, one may have a
\emph{``differential-geometric mechanism''}, in the sense of ADG,
\emph{different}, in character, \emph{from the classical} manner
of obtaining it (viz., via smooth manifolds, \emph{but}, see also
(4.1), along with (4.11) below). So it is, indeed quite useful
(yet, rather, necessary(!)) to make the following remarks. That
is,

\bino (4.1)\hfill
 \bmp
    even, if we take, as the \emph{base space of} the
    \emph{sheaves} involved, within the \emph{abstract context of}
    ADG, a (\emph{smooth}) \emph{manifold} $X$, in the standard
    sense of this term (cf. thus the ensuing two Subsections
    below), \emph{its r\^{o}le} (:\;as \emph{the} source of
    Calculus) \emph{is} actually \emph{transferred} now \emph{to
    the ``sheaf of coefficients''}, $\ca$. Yet, this is very
    \emph{organic}, since it is essentially \emph{we}(!), who
    \emph{make} the \emph{calculations}/experiments, based on our
    own \emph{``arithmetics''}, viz. for the case in hand, again,
    via the \emph{algebra sheaf} $\ca$.
 \emp

\bino Therefore,

\bino (4.2)\hfill
 \bmp
    the manifold $X$, as in (3.1) above, is just viewed, simply,
    as a particular \emph{topological space}, being, of course, by
    its very definition, \emph{paracompact} (\emph{Hausdorff});
    the latter condition is certainly, otherwise, very useful,
    indeed, when referring to \emph{cohomological} issues:
    \emph{sheaf cohomology} is, for that matter, apart from
    \emph{sheaf theory} itself, the other fundamental ingredient
    of ADG. Yet, it may still happen that the \emph{topology of}
    $X$ be chosen quite \emph{different from} the initial, viz.
    the \emph{standard} topology of the manifold $X$, i.e., the
    \emph{``locally euclidean''} one); see, for instance,
    \emph{``Sorkin's topology''} in Subsection 4.(b) below.
 \emp

\bino However, as we shall see, by the ensuing discussion, the
particular cases we look at in the sequel, \emph{do have}, so to
say, \emph{a}
\begin{align*}
    \text{\emph{newtonian spark}\,(!),}
    \tag{4.3}
\end{align*}
that is, something of a \emph{``starting point''}, that will
become better clear, by the subsequent rationale. Notwithstanding,
as we shall also realize, in that context,

\bino (4.4)\hfill
 \bmp
    \emph{this does not affect}, at all(!), \emph{the
    ``leibnizian''} character of the \emph{mechanism of} ADG,
 \emp

\bino as the latter is \emph{inherently afforded, by the} same two
particular \emph{subsequent examples} of the general theory.

The preceding certainly constitutes a \emph{fundamental} special
\emph{issue} of \emph{paramount importance}, indeed, for potential
applications; the same could still be worthwhile to be viewed
\emph{axiomatically}(!), as well, contributing thus to our
knowledge, as it concerns the \emph{whole character of the general
theory}. In this regard, see also our previous account thereof,
already in A.\;Mallios [17: Chapter IX; Section 5].

\bino $ $\hfill
 \bmp
    \hspace{0.5cm}{\bf Note 4.1.}$-$ By still referring to our
    previous issue in (4.3), as we shall see, by the ensuing
    examples the so-called therein \emph{``newtonian spark''} not
    only supplies the \emph{``structure sheaf''} $\ca$, by the
    \emph{``spark''} (fuse) of its \emph{``differential''
    mechanism}, but what is, in effect, herewith of a particular
    importance, is that \emph{one assures}, in that context, the validity of
    \emph{Poincar\'{e} Lemma}, indeed, of an \emph{extraordinary
    importance} of the whole mechanism of ADG. Thus, one can
    complete (4.3), by actually setting the \emph{equivalence};

    \bino (4.5)\hfill
    \bmi
        \emph{``newtonian spark'' $\Longleftrightarrow$
        Poincar\'{e} Lemma}.
    \emi
 \emp

\bino So here again one realizes the \emph{fitness of}

\bino (4.6)\hfill
 \bmp
    \emph{replacing of the} \emph{``geometric character'', locally}(!),
    \emph{of classical analysis}, \emph{by cohomological issues.}
\emp

\bino However, more on this we shall see in the pertinent places
below.

Thus, we come now to examine our first Example, pertaining to the
situation described by (4.3), (4.5) above, straightforwardly, by
the ensuing Subsection:

\bi {\bf 4.(a). Rosinger's algebra sheaf.}$-$ Here the
aforementioned already \emph{``newtonian spark''}, as in (4.3)
above, is nothing more, as we shall presently see, right below,
than \emph{the classical}
\begin{align*}
    \text{``$dx$''}
    \tag{4.7}
\end{align*}
of the \emph{standard theory of $\cc^\infty$-manifolds}. Thus,
\emph{the} above \emph{classical} ``$dx$'' is, for the case at
issue, prolonged, true, \emph{it is}, in point of fact,
\emph{``promoted''}(!), so to speak, \emph{to} an abstract,
\begin{align*}
    \text{``$\partial$''}
    \tag{4.8}
\end{align*}
\emph{in the sense of} ADG (see, for instance, (2.14) in the
preceding), defined now on an \emph{algebra sheaf} (Rosinger's),
containing the standard one $\cc^{\infty}_{X}$, viz. the
$\bc$-\emph{algebra sheaf} (of germs of $\bc$-valued smooth
functions [$\br$-valued functions could also be considered, of
course]) on a given manifold $X$. Indeed, we can still say, in
anticipation, that \emph{Rosinger's algebra sheaf} $\ca_{nd}$,
and, i\,n\, e\,x\,t\,e\,n\,s\,o\, $\ca_{foam}$ (see (4.20), (4.21)
in the sequel) \emph{contain much more than} $\cc^{\infty}_{X}$ of
the classical theory (cf. thus (4.14) below). We depict the above,
by the following diagram, whose notation will become more clear,
through the ensuing discussion. Thus, \emph{we have};

\bino (4.9)\hfill
 \begin{minipage}{10.5cm}
 \begin{picture}(300,120)%
 \put(45,10){\makebox{$\ca_{nd}$}}%
 \put(70,13){\vector(1,0){70}}
 \put(105,3){\makebox{$\scriptstyle{\partial}$}}
 \put(145,10){\makebox{$\Om^{1}_{nd}\equiv \Om^1$}}
 \put(55,95){\vector(0,-1){70}}
 \put(45,55){\makebox{${\bigcap}$}}
 \put(150,95){\vector(0,-1){70}}
 \put(139,55){\makebox{$\bigcap$}}
 \put(47,102){\makebox{$\cc^{\infty}_{X}$}}%
 \put(70,105){\vector(1,0){70}}
 \put(102,108){\makebox{$d$}}
 \put(145,102){\makebox{$\Om^{1}_{X}$}}
\end{picture}
\end{minipage}

\bino We proceed, by explaining the notation applied in (4.9);
thus,
\begin{align*}
    \ca_{nd} \equiv \ca,
    \tag{4.10}
\end{align*}
stands therein for \emph{Rosinger's algebra sheaf} a
$\bc$-\emph{algebra sheaf} on $X$, the latter space being, by
assumption, an \emph{open} subset \emph{of} $\br^k$. However,
since the whole theory is, in point of fact, of a \emph{local
nature}, one may consider, instead, $\br^k$ \emph{just locally},
that is, we can assume that $X$ is a \emph{smooth}
(:\;$\cc^\infty$-)\emph{manifold}. Notwithstanding, for
simplicity's sake, \emph{we adopt}, throughout, \emph{that}
\begin{align*}
    \text{$X$ is \emph{open in} $\br^k$}.
    \tag{4.11}
\end{align*}
Accordingly, $X$ \emph{being}, by its very definition, a
\emph{metrizable} space, one concludes, in particular, that

\bino (4.12)\hfill
 \bmp
    $X$ is a \emph{paracompact Hausdorff} (topological)
    \emph{space}.
 \emp

\bino See, for instance, J.\;Dugundji [6: p. 186, Theorem 5.3].

Now, \emph{Rosinger's algebra sheaf} $\ca \equiv \ca_{nd}$, as in
(4.10) above, is actually an appropriate (cf. (4.13) in the
sequel) \emph{quotient of a functional} (algebra) \emph{sheaf}:
thus, technically speaking, it is defined, as a \emph{quotient of
a functional} (algebra) \emph{presheaf}, the latter being proved,
in particular, to be a \emph{``complete''} one, therefore
(J.\;Leray), a \emph{sheaf}. Yet, the corresponding, in that
context, \emph{quotient algebras} are defined, modulo a suitable
(2-sided) \emph{ideal} (:\;\emph{``Rosinger's ideal''}), which is
essentially \emph{characterized, by} what we may consider, as
\emph{``Rosinger's asymptotic vanishing condition''}; in
particular, the latter is defined, via a

\bino (4.13)\hfill
 \bmp
    \emph{closed nowhere dense} (hence, the subindex ``nd'',
    appeared in (4.10)) \emph{subset} $\Ga$ \emph{of} $X$, the
    same ideal consisting thus of those \emph{functions}/elemenets
    of the (local section) algebras concerned, that \emph{vanish
    ``eventually''} (w.r.t. a parameter involved, a natural
    number, index) \emph{on any relatively compact} subset
    \emph{of the complement of $\Ga$}.
 \emp

\bino Concerning the precise definition of the preceding, we refer
to A.\;Mallios [17: Chapt. IX; Section 5], or even (:\;to
A.\;Mallios-E.E.\;Rosinger [22: p. 236; (2.2)]. Yet, by further
looking at the same sheaf (4.10), as above, and also complementing
the information we have through (4.9), we still note that we
actually get, by the very definition of (4.10) (cf., for instance,
E.E.\;Rosinger [29: p. 8; (1.2.15), (1.2.16), along with p. 367,
(2)]),
\begin{align*}
    \cc^{\infty}_{X}\varsubsetneqq \mathfrak{D}'_{X} \subseteq \ca
    \equiv \ca_{nd}.
    \tag{4.14}
\end{align*}
Here the middle term in (4.14) denotes the \emph{sheaf} (of germs)
\emph{of Schwartz distributions} on $X$, viewed, as a
$\bc$-\emph{vector space sheaf} on $X$ (loc. cit., (5.19)).

On the other hand, the \emph{``basic differential operator''}
\begin{align*}
    \partial :\ca \;\lto \;\Om^1 ,
    \tag{4.15}
\end{align*}
that one has to define, according to the general theory of ADG,
see, for instance, (2.14) in the preceding, or even in A.\;Mallios
[13: Chapt. VI; Section 1], is here provided by the presence of
the first member in (4.14), that is, \emph{locally}, by that one
of a ($\bc$-)\emph{algebra} of the form
\begin{align*}
    \cc^\infty (U), \ \text{ with } \ U \ \text{\emph{open} in } \
    X\subseteq \br^k ,
    \tag{4.16}
\end{align*}
(see also (4.11)), that virtually constitutes, within the present
context, the \emph{``newtonian spark''}, hinted at in (4.3). Thus,
the \emph{basic differential} $\partial$, as in (4.15), in now
defined \emph{coordinate-wise}, along the classical patterns,
since the \emph{basic constituents of} the \emph{Rosinger's
algebra} (\emph{pre})\emph{sheaf} are (local sections of)
\emph{cartesian product algebras} of the form,
\begin{align*}
    (\cc^\infty (U))^\bn ,
    \tag{4.17}
\end{align*}
with $U$, as in (4.16), which then are \emph{``quotiented'',
according to} (4.11). Of course, the previously coordinate-wise
(\emph{classically}!) defined \emph{differential} passes to the
quotient. For technical details see A.\;Mallios [17: Chapt. IX;
Subsection 5.(b)], or even to A.\;Mallios-E.E.\;Rosinger [22].
Thus, the overall moral, that is here, concluded according to the
\emph{general principles of} ADG, is the following;

\bino (4.18)\hfill
 \bmp
    Starting \emph{from any} basic \emph{``differential triad''},
    in the sense of ADG (even a classical one, as e.g. a
    \emph{``locally euclidean} one, this is the case, herewith,
    \emph{we} can then \emph{perform any} (functorial)
    \emph{operation, provided within the category of differential
    triads}, to get thus at a new one [occasionally, more
    useful/flexible than the initially given one!].
 \emp

\bino Thus, by referring, in particular, to \emph{Rosinger's
algebra sheaf,} as above, \emph{and the associated} with it
\emph{differential triad}, we remark that, in view of (4.18), what
we actually consider, in that context, is:

i) to take a \emph{denumerable cartesian product} of the standard
(:\;\emph{newtonian-cartesian}) \emph{differential triad}
\begin{align*}
    (\cc^{\infty}_{X}, d,\Om^1 ),
    \tag{4.19}
\end{align*}
as well as,

ii) to take, in particular, \emph{a} pertinent \emph{quotient} of
the above, modulo \emph{Rosinger's ideal}, as indicated by (4.13).

In this connection, the aforesaid \emph{categorical treatment of}
ADG, has been occasionally considered already in A.\;Mallios [13:
Chapt. VI; Sections 5, 6], as well as, in [17: Chapt. I; Section
5.(e), 5.(f): \emph{``pull-back''} functor]; yet, an analogous
fuller and systematic \emph{categorical study of differential
triads} has been recently supplied by the relevant work of
M.\;Papatriantafillou [25], [26], [27].

On the other hand, one gets at an \emph{immense generalization} of
the above, by considering, in place of $\ca_{nd}$, what we may
call a \emph{Rosinger's multi-foam algebra sheaf}, along with the
associated \emph{differential triad},
\begin{align*}
    (B_{\La ,J}, \partial, \Om^1 );
    \tag{4.20}
\end{align*}
here the space $X$, base of the sheaves concerned, is still given
by (4.11), while the sheaf on $X$ appeared in the first member of
(4.20) is again a pertinent \emph{quotient} of the ($\bc$-)algebra
\begin{align*}
    \cc^\infty (X)^\La ,
    \tag{4.21}
\end{align*}
with $\La$ an \emph{upwards directed set}, modulo an analogously
defined (2-sided) \emph{ideal} of the same algebra, with respect
to a given (upwards) \emph{directed family $J$} of
\emph{``residual''} subsets of $X$, the
\emph{``singularity-sets''} of $X$ (viz. those $A\subseteq X$,
with $\overline{\complement A} =X$), the applied terminology,
herewith, being hinted at potential physical applications: See
A.\;Mallios-E.E.\;Rosinger [23], as well as, A.\;Mallios [17:
Chapt. IX; Section 6]. Of course, the \emph{singularity-sets}, as
above, \emph{generalize} the notion of \emph{nowhere dense sets},
considered by (4.13) in the preceding. Hence, the \emph{increase
of the types of ``singularities'', one can cope with}, in the
framework of ADG, as explained in the foregoing.

Now, \emph{the} same \emph{moral}, that dominates our previous
comments \emph{in} (4.18), is, in point of fact, as we shall
presently see in the sequel, the \emph{prevalent point of view}
also \emph{in} the ensuing example, referring to another
\emph{potential application} of the very technique of ADG
\emph{in} problems of \emph{quantum gravity}.

\bi {\bf 4.(b). Finitary incidence algebra sheaves.}$-$ Similarly
to the preceding Example 4.(a), here too, as already said, for
that matter, one starts again from a \emph{smooth}
(:\;$\cc^\infty$-)\emph{manifold} $X$, that still, for
simplicity's sake, we assume that it is just an \emph{open} subset
\emph{of} the \emph{euclidean space} $\br^k$ (see (4.11)).
However, as we shall see, \emph{this} important (:\;\emph{very
restrictive}(!), otherwise) \emph{hypothesis} will finally be
\emph{used, only}(!) \emph{in connection with} (4.5)(!), as that
was the case in the foregoing, as well: Thus,

\bino (4.22)\hfill
 \bmp
    \emph{no ``global use/presence'' of} the \emph{euclidean or
    even locally euclidean space is made}, at all(!).
 \emp

\bino This \emph{important fact}, indeed, ensures actually, the
associated method, as it concerns, at least, its
\emph{differential-geometric nature}, its \emph{potential
versatility}.

Now, following R.\;Sorkin [32], one chooses the \emph{locally
finite open coverings} of $X$ (recall that the latter space is
here also \emph{paracompact Hausdorff}, see e.g. (4.12) in the
preceding), while one further considers on the \emph{set} $X$ the
\emph{topology generated by} such \emph{locally finite open
coverings of} $X$, as above. In this connection we also recall,
for occasional use, in relation with ADG, that;

\bino (4.23)\hfill
 \bmp
    the \emph{local frames} of a given \emph{vector sheaf} on a
    \emph{paracompact} (Hausdorff) space $X$ constitute a
    \emph{cofinal subset of} the \emph{locally finite open
    coverings} of $X$.
 \emp

\bino See [VS: Chapt. IV; p. 325, (8.42), along with Chapt. II: p.
127; (4.9)].\rl

Now, the previous topological spaces, that are associated with
locally finite open coverings of $X$, are further endowed, \`{a}
la Sorkin (loc. cit.), with appropriate partial orders, becoming
thus \emph{``posets''}, alias, \emph{``fintoposets''}, in the
terminology of I.\;Raptis [28] (see also A.\;Mallios-I.\;Raptis
[29]). On the other hand, these toposets are further suitably
associated with certain finite-dimensional associative
(non-abelian) linear $\bc$-algebras, the so-called
\emph{``incidence Rota algebras''} (loc. cit.). The same algebras
are further sheafified, the resulting sheaves leading finally to
appropriate \emph{``differential triads''}, in the sense, of
course, that this notion is used by ADG (see [VS: Vol. II]). Here
again, as it also was the case in our previous example in
Subsection 4.(a) above, it is of a crucial significance the

\bino (4.24)\hfill
 \bmp
    \emph{possibility of using} the item connected with what we
    have called in \emph{the} preceding, \emph{``newtonian
    spark''} (cf. thus (4.5)).
 \emp

\bino As it was pointed out therein, the latter issue is the
\emph{``source''} of the \emph{``differential mechanism''}, that
one is supplied with, yet within the present context too,

\bino (4.25)\hfill
 \bmp
    without employing, in effect, the euclidean, or even locally
    euclidean nature of the origin of that particular
    \emph{``spark''}, in the way, at least, we are used to do it in
    the classical theory, thus far!
 \emp

\bino However, for the technical details thereof, we refer to the
relevant work of A.\;Mallios-I.\;Raptis [20], along with that one
of the same authors in [21]. We have thus herewith still another
realization of the fact, being, in point of fact, a
\emph{fundamental moral of} ADG (see also A.\;Mallios [14]), that;

\bino (4.26)\hfill
 \bmp
    when we try to apply (\emph{differential}) \emph{geometrical
    methods}, more so \emph{in the quantum deep}, it seems
    \emph{more natural} to \emph{apply an analytic} (:\;algebraic)
    \emph{way} (with symbols --recall here, for instance,
    \emph{``Feynman diagrams''}-- viz. a \emph{``Leibnizian''} manner
    of looking at the things, in focus), \emph{not that one of the
    standard theory} (:\;\emph{``spatial-newtonian''}).
 \emp

\bino Yet, what actually leads to the same thing,

\bino (4.27)\hfill
 \bmp
    it is quite natural to try to \emph{concoct}, at each
    particular case, under con\-si\-deration, \emph{the appropriate
    ``differential geometric''-machinery} (viz.
    \emph{``differential triad''}), to cope with the problem at
    issue.
 \emp

\bino In toto, we could also mention herewith, a \emph{basic moral
of} ADG, in what actually concerns \emph{Quantum Field Theory}.
That is,

\bino (4.28)\hfill
 \bmp
    we should \emph{not relate} any (quantum) \emph{field theory
    with} the existence of an a\,d\, h\,o\,c\, given
    \emph{``continuum''} (:\;``space-time manifold'', whatsoever);
    this, of course, to the extent, at least, that we wish to
    apply therein (classical) differential geometry (CDG), since,
    in that context, the preponderant and really instrumental
    issue is, in effect, \emph{the} relevant
    (differential-geometric) \emph{technique} and \emph{not}(!)
    \emph{the underlying space}.
 \emp

\bino So, in other words, it is important to afford, in that
context, a \emph{``differential-geometric'' machinery},
irrespective of the way the latter might have been displayed (cf.,
for instance, the preceding two examples), while, in any case,
\emph{this particular way, ``spatial'', or not} (loc.\;cit.),
\emph{should not intervene in the whole process}, this being
es\-pe\-cially significant, when referred to the \emph{quantum
r\'{e}gime} (see also the relevant comments already in (1.19) in
the preceding).

Indeed, in this regard, we can still remark that,

\bino (4.29)\hfill \bmp as it concerns the \emph{``infinitely
small''} (Feynman), \emph{the} (differential) \emph{``geometry''},
in the way, at least, that we use to look at it (viz. in the
\emph{``newtonian-cartesian''} one), \emph{is no more valid}(!),
since the same --namely, the \emph{``geometry''} becomes --in
point of fact, appears to us --in that deep, \emph{more
``physical''}(!), as it always is, for that matter, viz.
\emph{``relational''} (:\;algebraic-analytic)! \emp

\bino Exactly at this point, we might also recall the quite
relevant remarks here of D.R.\;Fin\-kel\-stein [8: p. 155], in
that (emphasis below is ours);

\bino (4.30)\hfill
 \bmp
    \emph{``Physics was dominated by the Cartesian epistemology
    untill the quantum theory.''}
 \emp

\bino Relate the above with our previous considerations in
Scholium 2.1 in the preceding. Yet, as a further illumination of
the point of view of the whole \emph{formalism of} ADG, we have to
point out/clarify, herewith, once more, two fundamental issues of
the aforesaid perspective, that also provide a potential
outstanding application of the above formalism to ever present
problems, thus far, of \emph{quantum gravity}. That is, we have to
note, in that context, that:

i) One can employ ADG, \emph{as a} (differential)
\emph{``geometry''}, in the classical sense of the latter term,
\emph{even in the quantum deep}(!), provided, of course we accept
the following correspondence/\emph{``identification''}
(:\;axiomatic),
\begin{align*}
    \textit{fields $\longleftrightarrow$ vector sheaves},
    \tag{4.31}
\end{align*}
that is, in other words what we have already called elsewhere
\emph{``Selesnick's correspondence''} (see, for instance,
A.\;Mallios [17: Chapt. II], for a detailed account of this
subject matter).

ii) \emph{The same ``geometry''}, as above (viz. always, within
the framework of ADG), \emph{can still be construed, as} a
\emph{``dynamical variable''}, as well (see (2.6), in conjunction
with (2.13), as well as, with (2.12)).

On the other hand, another \emph{technical issue}, that should
also be pointed out in this regard, is that, \emph{it, very
likely, seems} that;

\bino (4.32)\hfill
 \bmp
    \emph{there is no} actually \emph{need} to
    \begin{align*}
     \textit{``quantize analysis''},\tag{4.32.1}
    \end{align*}
    as it concerns, in particular, its \emph{topological-linear
    character} (this being the \emph{source of the Calculus}),
    since the \emph{inherent/deeper nature} of the
    same (:\;of the \emph{``analysis''}), namely, the \emph{``algebraic''}, or
    even \emph{the}, so to say, \emph{``leibnizian'' one, is
    already}, viz., by its very definition, \emph{``quantized''}!
 \emp

\bino Yet, by further commenting on our last claim, as above, we
still note that;

\bino (4.33)\hfill
 \bmp
    \emph{there is no}, in effect, according to the same
    definitions, any \emph{``infinite'' in} (pure) \emph{algebra}!
 \emp

\bino So \emph{it is}, therefore, \emph{in ``geometry''/topology}
(viz. in the so-called, \emph{``infinite''}(!), a consequence, in
fact, of the latter perspective), that \emph{we are}, actually,
\emph{entangled}, when confronting with the \emph{quantum deep}
(:\;``small distances''). Consequently, our systematic endeavor,
up to this day, in one way or another, to succeed in getting an
appropriate \emph{``algebraization''} of the whole scenario!

\bi \bi {\bf 5. Scholium} (:\;more on the {\bf ``newtonian
spark''}).$-$ We usually \emph{curve a linear structure}, by
\emph{``localizing''} it (manifolds); in point of fact, this is a
quite \emph{general device}, referring, irrespective of the
dimension (finite or infinite), to the (\emph{topological})
\emph{vector space-model} of our (cartesian) \emph{``geometry''}.
In the case of Analysis, an extraordinary issue, in that context,
is that the classical Calculus that traditionally was hospitalized
in (even, emanated from) topological vector space-structures
(:\;euclidean spaces) still survived after this transport, a sine
qua non, of course, of the justification, for that matter, of the
previous movement, suggested, indeed, by particular important
applications. Notwithstanding, a \emph{fundamental moral} of the
whole issue \emph{of} ADG is that;

\bino (5.1)\hfill
 \bmp
    the real \emph{corner-stone} of the previous total enterprize
    is, in effect, what we have already called in the foregoing,
    the \emph{``newtonian spark''}, in that context, a fact that
    might also be paralleled with the famous \emph{archimedean
    demand}, for a pedestal (:\;``$\Delta\acute{o}\varsigma~\mu o\acute{\iota}~\pi\tilde{\alpha}~\sigma\tau\tilde{\omega}~\kappa\alpha\acute{\iota}~
    \tau\acute{\alpha}\nu~\gamma\tilde{\alpha}\nu~\kappa\iota\nu\acute{\eta}\sigma\omega$---{\it ``give me somewhere to stand and I shall move the
    earth''}).
 \emp

\bino That is, in other words, \emph{following} now
\emph{Leibniz}, in what actually concerns (classical) differential
geometry (CDG), what one virtually needs is to provide (according
to ADG) the appropriate, concerning the particular problem, at
issue, \emph{``differential-geometric mechanism''}(!).

Furthermore, what is here of a particular significance, having
also important potential applications (even, very likely(!), in
\emph{quantum gravity} too), is that:

\bino (5.2)\hfill
 \bmp
    \emph{the} aforesaid \emph{``differential-geometric
    mechanism''}, in the sense of ADG, \emph{does not} actually
    \emph{depend}, at all(!), \emph{on any space}, as it was the
    case, so far, for the classical theory (CDG), the same
    mechanism \emph{referred} now \emph{directly to the}
    (``geometric'') \emph{objects}, that live on the
    \emph{``space''}.
\emp

\bino Indeed, the latter issue in the above remarks, as in (5.2),
is, most likely, what already Leibniz, at his time, was looking
for! (See, for instance, N.\;Bourbaki [4: Chapt. I; Note
historique, p. 161, ft. 1], or even A.\;Mallios [14: (2.1), along
with comments following it]).

Thus, by further commenting on (5.2), we can still say, based also
on our previous considerations in A.\;Mallios [15], that;

\bino (5.3)\hfill
 \bmp
    What one actually perceives appears to be the
    \emph{``sheafification''} of a \emph{``local
    aspect/information''} pertaining to the particular subject
    matter in focus. Besides,

    \bino (5.3.1)\hfill
    \bmi
        the way we get a \emph{``local information''}, may, in
        principle, be \emph{entirely different, in character},
        from the \emph{mechanism} (:\;inherent
        law--\emph{``physical''/relational procedure}), \emph{which
        governs} (hence, the manner too, we should essentially
        employ the aforesaid \emph{``sheafification''}, viz. the
        \emph{global aspect of}) \emph{that local information}.
    \emi
\emp

\bino The above explains too what one essentially encounters, in
connection with what we have called in the preceding
\emph{``newtonian spark''}.

Now, the replacement of a \emph{``field''}
\begin{align*}
    (\ce, D)
    \tag{5.4}
\end{align*}

\bino (cf. (2.15)), by its corresponding \emph{``Heisenberg}
(:\;``matrix'') \emph{picture''}, viz. by the \emph{``field''}
\begin{align*}
    (\ce nd \ce ,D_{\ce nd \ce}),\tag{5.5}
\end{align*}
(see A.\;Mallios [14: (9.20)], along with A.\;Mallios [17: Chapt.
VII; (5.8), (5.11)]), hence, via its \emph{``principal sheaf''}
version,
\begin{align*}
    (\ca ut \ce, D_{\ce nd \ce}|_{\ca ut \ce}),\tag{5.6}
\end{align*}
as well, may still be viewed, as being in accord with the
``impossibility of having a \emph{``relativistic quantum field''},
defined at a point''(!); see, for instance, N.N.\;Bogolubov et al.
[4: p. 282, \S 10.4, p. 283, Theorem (Wightman) 10.6].

On the other hand, (5.3.1), as above, might also be construed, as
another effectuation of the classical \emph{``local
commutativity''}, or \emph{``microscopic causality''}
\emph{``microcausality''} yet, \emph{``principle of relativistic
microcausality''}, or even \emph{``Einstein's locality''}.

On the other hand, by further meditating, a bit more, on our
previous scholium in (5.3.1), we can actually reformulate it, as
well, by remarking, in particular, that:

\bino (5.7)\hfill
 \bmp
    \emph{the deeper} (algebraic) \emph{mechanism} that might be
    \emph{inherent in} (:\;esoteric \emph{of}) \emph{a} given
    \emph{local information} (alias, of given \emph{local data}),
    \emph{may}, in general, \emph{be}, quite well,
    \emph{independent of the way, one} has \emph{drawn this
    information} (:\;\emph{the local data}, concerned).
\emp

\bino Yet, in connection with the above remarks in (5.7) and our
issue in (4.3), one may recall, in this regard, Wittgenstein's
motto [37: p. 74; 6.54], in that;

\bino (5.8)\hfill
 \bmp
    ``...\;[one must]... \emph{throw away the ladder after he has
    climbed up it}.''
 \emp

\bino Now, as a fundamental spinoff of the above, one can still
conceive, for instance, in that context, the classical
(\emph{Machian}) perspective of an
\begin{align*}
    \text{\emph{``action at a distance''}(!)}.
    \tag{5.9}
\end{align*}

\bi \bi {\bf 6.\;Concluding remarks (the ``continuum'')}$-$ The
purpose of this final section is to make clear, once more, that:

\bino (6.1)\hfill
 \bmp
 the notion of the \emph{``continuum''}, as a \emph{``foundational
 element'' is not} actually \emph{the case}, when \emph{physically
 speaking}, at least\,(!) (and not only\,(!), see e.g. (6.5) in
 the sequel).
\emp

\bino Now, the inverted commas put on the word continuum, as
above, refer, of course, to the way \emph{we} usually understand
that notion in the familiar terminology of the classical theory,
where, in point of fact, \emph{we} wish to ascribe to it a
physical substance, that is, equivalently, to endow it with a
physical meaning. And just hear one has \emph{the crux of the
problem}: That is,

\bino (6.2)\hfill
 \bmp
    we are actually influenced by \emph{our mathematical
    terminology--conception}, in what virtually concerns the word
    \emph{``continuum''}, i.e., the \emph{``cartesian''}, in point
    of fact, perspective of the so-called \emph{``space-time''}.
 \emp

\bino Thus, in other words, we make the following
\emph{identifications}:

\bino (6.3)\hfill
 \bmp
    \emph{``physical space''} $\longleftrightarrow$
    \emph{mathematical ``space''/``continuum''}, viz. some
    $\br^n$, as a (finite dimensional) \emph{topological vector
    space}.
 \emp

\bino However, it is exactly \emph{the above identifications},
that \emph{is} really \emph{the source of the problems}: Indeed,
as we have already remarked in other places (cf., for instance,
A.\;Mallios [14: (1.4), or even (3.1)]),

\bino (6.4)\hfill
 \bmp
    \emph{``physical space''} is \emph{what} virtually
    \emph{constitutes it}, that is, in other words, what we may
    call, \`{a} la Leibniz, the \emph{``geometrical objects''}
    themselves, that make up, what in effect, we perceive, as
    \emph{``space'', in the large}, as well as, \emph{in the
    small}.
 \emp

\bino Therefore, in that respect, the substance of the
\emph{``physical space''}, as above, is thus
\emph{discrete/granular}, hence not at all corresponding to
something \emph{``continuous''}, viz. \emph{not-discrete}, when
physically/conceptually speaking. On the other hand, when
\emph{mathematically speaking}, a \emph{set} is already, by its
very definition, being thus \emph{``point-wise determined''},
absolutely \emph{``discrete''}, in character!

Thus, by referring to \emph{the mathematical notion of the
``continuum'', as an} $\br^n$, $n\in \bn$,

\bino (6.5)\hfill
 \bmp
    we note that the so-called \emph{``continuum'' is},
    technically speaking, viz. as a \emph{mathematical term, our}
    own \emph{definition} of an $\br^n$ $(n\in \bn)$, as already
    said, viewed herewith not just, as a \emph{discrete set}, as
    it actually is, for that matter, but now, as a
    \emph{topological} (\emph{vector}) \emph{space}, this
    particular (mathematical) \emph{``structure''} on (the set)
    $\br^n$ being also the \emph{source of the} (\emph{newtonian})
    \emph{Calculus}!
 \emp

\bino In this connection, we are thus influenced, by our own
\emph{mathematical experience} of the concept of the
\emph{``continuum''}, in the way we defined it, as above, that is,
as a particular \emph{finite dimensional} (\emph{Hausdorff})
\emph{topological vector space}, a point of view that we also
attribute, in turn, to what we actually perceive, as a
\emph{``physical space''}, this being further construed, as
another ``continuum'', this time, however, as a physical one\,(!),
based rather on a \emph{``dynamically/kinematiccaly''} ascribed
description of the (physical) world; alas, something here \emph{in
complete conflict with} our actual (:\;experimental) experience,
as it virtually concerns, at least, \emph{the quantum r\'{e}gime}
(see also, for instance, (4.30) in the preceding).

Now, in this context, the previous items;

\bino (6.6)\hfill
 \bmp
    \emph{``dynamical-kinematical description''} of the (physical)
    world, \emph{differential equations-theoretic point of view,
    Calculus}, and \emph{``space-time continuum''} are, in effect,
    \emph{intimately related} and, in point of fact,
    \emph{tautosemous}, in substance.
 \emp

\bino Strictly speaking, as a matter of fact,

\bino (6.7)\hfill
 \bmp
    \emph{Calculus} is the source/cause of the first two items, as
    above, while, in turn, the same (Calculus) is \emph{the
    spin-off}, as already said, \emph{of the newtonian-cartesian},
    so far, \emph{definition of the ``space''}, that is, of the
    so-called \emph{``geometrical'' perception} of it, yet,
    \emph{the outcome of} the same \emph{``space-time
    continuum''}$\equiv \br^n$ (cf. (6.5)).
 \emp

\bino On the other hand, the above \emph{differential part of the
Calculus}, viz. \emph{``differentiation of functions''}, in
principle, presupposes \emph{``good}(-- (:\;smooth)
differentiable)--\emph{functions''}, something that essentially
depends on the \emph{``local behavior''} of the functions
concerned; hence, a fact that directly refers to the \emph{local
nature of the domain of definition of the} same \emph{functions},
that is, to the \emph{local structure} (:\;\emph{``geometry''}) of
the \emph{``euclidean space''}, $\br^n$, itself. Consequently,

\bino (6.8)\hfill
 \bmp
    we are actually \emph{entangled with the way} the
    \emph{differential calculus} (:\;\emph{``differentiation''},
    as a mechanism) is supplied (cf. (6.5)), therefore, the
    \emph{type of the ``differentiable'' functions} that are
    \emph{thereby involved}, or, in other words, that are
    \emph{``locally'' defined} on that particular \emph{``space''}
    (\emph{extremely important}, as well as,
    \emph{effective}\,(!), anyhow, concerning the classical
    theory).
 \emp

\bino However, the \emph{applications of} the same
\emph{differential calculus}, as above, \emph{in} the domain of
(classical) \emph{differential geometry}, as a means of study
(:\;working instrument) in that particular discipline, namely,
that what we have already considered in the preceding, a
\emph{``differential-geometric machinery''}, yet, in other words,
a \emph{``geometrical calculus''}, \`{a} la Leibniz (see 1.17)),
refers, in point of fact, to the very \emph{``geometrical
objects''} (Leibniz, loc. cit.), the same being actually (Leibniz,
ibid., Riemann, see [14: (1.3)]) \emph{independent of any
``space''}, in the sense, at least, of (6.5), as above!

On the other hand, it is reasonable to think that,

\bino (6.9)\hfill
 \bmp
    the very character of what we may call

    \bino (6.9.1)\hfill
    \bmi
        \emph{``physical space''} (see also (1.1)) \emph{is}, in
        principle, \emph{the same}, both \emph{in the large, as
        well as, in the small}.
    \emi

    \bino We are thus led to a dissonance, by applying our usual
    \emph{classical re\-presentation of the physical space} (in the
    large), \emph{as an} $\br^n$, irrespective, of course, of the
    tremendous success, thus far, of the latter perspective, when
    \emph{realizing}, on the other hand (see also, for instance,
    (2.2), as well as, (4.30) in the preceding), \emph{that the
    same} (physical) \emph{``space'' is} virtually \emph{quite
    different from what we are confronted with}, when looking
    \emph{at the quantum r\'{e}gime}, as it concerns the aforesaid
    classical perspective; see also A.\;Mallios [14: (8.10), along
    with (8.11)].
 \emp

\bino Consequently, the appeared inconveniences
(:\;\emph{``singularities''}), regarding, of course,
\emph{applications of classical differential geometry}, as a means
of study, in that context, of the \emph{physical space/geometry''
in the small}, that is, to say, \emph{physical laws/``fields''}
(see also loc. cit., (3.21.1)) \emph{at the ``quantum
resolution''}.

Thus, the \emph{``physical space''}, as a whole, yet, according to
recent advances in theoretical physics, concerning, in particular,
the \emph{quantum deep}, does not seem to be the usual
\emph{``space-time'' manifold}, in the sense of the
\emph{classical differential geometry}--theory of \emph{smooth}
(:\;$\cc^\infty$-)\emph{manifolds}. Indeed, it appears that we
have therein,

\bino (6.10)\hfill
 \bmp
    something \emph{foamy, very singular}, or even something like
    what we may call, a \emph{``singularity manifold''}, to refer,
    in that respect, to a rather recent utterance of R.\;Penrose,
    pertaining to a \emph{``true theory of quantum}
%
    \emph{gravity''}, by
    replacing the \emph{``present concept of spacetime at a
    singularity''}. (See also, for instance, A.\;Mallios [14:
    (10.8), along with the subsequent discussion therein]).
 \emp

\bino Thus, concerning the \emph{``infinitely small''}, or else
\emph{``quantum resolution''},

\bino (6.11)\hfill
 \bmp
    the \emph{``geometry''}, in the way we usually look at it
    (viz. in the \emph{``newtonian-cartesian''} manner), is
    \emph{no more valid}\,(!), in that context, since, at that
    deep, the same \emph{becomes} thus, even to our senses\,(!),
    more \emph{``physical''}, as, in point of fact, it always is,
    for that matter (see also (1.1) in the preceding, along with
    (6.9) above), viz. \emph{``relational''}
    (:\;algebraic-analytic); the latter aspect is \emph{still a
    fundamental issue}, in effect, of our experience, thus far,
    \emph{from} ADG, as  well!
 \emp

 \bino Indeed, as already advocated in several places in the
 preceding, working within the context of ADG, we are able to look
 at fundamental concepts of physics, as, for instance,
 particles/fields (see also A.\;Mallios [16: (3.2)]) etc, without
 being compelled to stick to such \emph{``technical'' notions}, as
 e.g. \emph{``space-time''}\,(!). In this regard, see also, for
 instance, (1.8) in the preceding, yet, loc. cit. (1.6).

 Yet, in this connection, we are thus very likely, led to conclude
 that the entanglement of the \emph{``manifold''} perspective in
 nowadays physics, especially, \emph{in the quantum domain}, is to
 be attributed, in effect, to the \emph{relation of the former
 with the notion of ``differential''}; indeed, the latter is the
 main function, that is actually applied in our relevant
 rationale, in that context, while finally, as an upshot of the
 classical theory (:\;differential geometry), we are usually of
 the opinion that \emph{the} same \emph{manifold} concept \emph{is
 thus the unique}(!) \emph{source of} the notion of a
 \emph{``derivative''}, covariant or not! \emph{Therefore, the
 significance,} hereupon \emph{of} the new proposal, as this is, provided, by
 \emph{ADG}: That is, once more, we realize that

 \bino (6.12)\hfill
  \bmp
    \emph{to have a connection, we do not} actually \emph{need a
    ``manifold''}, even if we momentarily borrow from such a
    concept the \emph{``infinitesimal} (:\;\emph{``newtonian''})
    \emph{spark}! The latter constitutes, in point of fact, the
    \emph{central moral} of the entire study \emph{of ADG}.
  \emp

 \bino Thus, by looking at the \emph{standard question} (cf., for
 instance, R.W.\;Sharpe [30: p. 2, ft. 2]),

 \bino (6.13)\hfill
  \bmp
    \emph{``what geometry on a manifold supports physics?''},
  \emp

\bino we can combine it now with the \emph{fundamental moral of
ADG}
 (see e.g. (6.12), as above, along with A.\;Mallios [14: (1.2),
 (3.21.2)]), that, in point of fact,

 \bino (6.14)\hfill
  \bmp
    \emph{differential geometry} means, in effect,
    \emph{connection}.
  \emp

 \bino Consequently, blending the previous two aspects, as in
 (6.13) and (6.14), we are thus led to a \emph{response to}
 (6.13), in the sense of affording a \emph{pertinent choice of}
 ``$\ca$'', more precisely speaking, the associated with it
 \emph{``differential triad''} and the concomitant
 \emph{``differential-geometric mechanism''}, \`{a} la ADG,
 suitable to the particular problem at issue; see thus, for
 instance, Subsections 4.\,(a), 4.\,(b) in the preceding, along
 with A.\;Mallios [12:\;p.\;174; concluding remarks]. In this
 regard, see also the latest re\-le\-vant account in
 A.\;Mallios--I.\;Raptis [21].

 \bi {\bf 6.(a). ADG vis-\`{a}-vis a Unified Field Theory.}$-$
 The quite ambitious(!) title of the present Subsection is rooted,
 in point of fact, on our previous \emph{comments in} (6.9.1) and
 on the very essence of the point of view of the same \emph{Abstract
 Differential Geometry} (ADG), as the latter can be applied, in
 that context, in conjunction, for instance, with \emph{Rosinger's
 theory of ``generalized functions''}: The technical part of the
 aforesaid scheme, hinted at herewith, has been already expounded
 in A.\;Mallios-E.E.\;Rosinger [22], [23]; in this connection, see
 also our previous discussion in Subsection 4.(a) in the
 foregoing, along with A.\;Mallios [16: (5.20), (5.21)], as well
 as, [14: Sections 6, 8; see, in particular, (8.8) therein, or
 even (8.11), yet, Section 10]. Moreover, cf. also A.\;Mallios
 [17: Chapt. IX; Sections 5, 6, along with Section 10 therein, see
 e.g. (10.29)].

 Now, as already said, the preceding just hint at a
 \emph{potential confrontation} with the second issue in the title
 of this Subsection, through the machinery of ADG, that is, to
 say, in terms of the \emph{techniques of the classical
 differential geometry}, being, however, \emph{freed now from} the
 ever disturbing/pestilential \emph{``singularities''}, and the
 like, of the classical approach to the problem at issue. Of
 course, this is due here, as already explained, throughout the
 preceding, to the \emph{absence}, according to ADG, \emph{of any
 supporting ``space'', that would also exclusively supply}
 (:\;generate) \emph{the ``differential-geometric'' machinery
 employed}, in that context (see the relevant citations, as
 before), a situation inherent, in effect, in the classical theory
 (CDG; cf. the previous Section 5, along with the concluding remarks
 above, preceding the present Subsection).

\end{document}